# Quantum Oscillations in iron-doped topological insulator $Sb_2Te_3$ crystals

Weiyao Zhao[a], David Cortie[a,b] Lei Chen[a], Zhi Li[a,b], Zengji Yue[a,b], and Xiaolin Wang[a,b*]

[a] *Institute for Superconducting and Electronic Materials, Australian Institute for Innovative Materials, University of Wollongong, NSW 2500, Australia*

[b] *ARC Centre of Excellence in Future Low-Energy Electronics Technologies FLEET*

**Abstract:** We investigated the magnetotransport properties of Fe-doped topological insulator $Sb_{1.96}Fe_{0.04}Te_3$ single crystals. With doping, the band structure changes significantly and multiple Fermi pockets become evident in the Shubnikov-de Haas oscillations, in contrast to the single frequency detected for pure $Sb_2Te_3$. Using complementary density functional theory calculations, we identify an additional bulk hole pocket introduced at the Γ point which originates from the chemical distortion associated with the Fe-dopant. Experimentally, both doped and undoped samples are hole-carrier dominated, however, Fe doping also reduces the carrier density and mobility. The angle dependent quantum oscillations were analyzed to characterize the complex Fermi surface and isolate the dimensionality of each SdH feature. Among the components, at least two pockets possess 2D-like behavior according to their rotational dependence. These results indicate a complex interplay of hybridized dopant bands with the bulk and surface topological electronic structure.

*Corresponding author: xiaolin@uow.edu.au

# 1. Introduction

Topological insulators (TIs) are materials with a symmetry-protected non-trivial electronic structure which can yield an insulating bulk with conducting boundaries. The emergent Dirac surface states are attractive for electronic applications and potentially host a range of fascinating phenomena including the quantum spin Hall effect, topological magnetoelectric effect, magnetic monopole images and Majorana fermions.[1-4] In 3D topological insulators such as the $(Bi,Sb)_2Te_3$ family, the surface electronic structure is entangled with the bulk electronic structure, and consequently, both aspects need to be understood at the fundamental level. Currently, there are a number of unresolved questions concerning the effect of transition metal doping in the $(Bi,Sb)Te_3$ family and this is related to one of the most fascinating transport properties in the Tis, namely the quantum anomalous Hall effect (QAHE). Shortly after it was predicted [5], the QAHE was experimentally realized in doped $(Bi,Sb)_2Te_3$[6] thin films. The Hall resistance reaches the predicted quantized value of $h/e^2$, accompanied by a considerable drop in the longitudinal resistance[6], which would significantly reduce the power consumption in electronic devices. However, the onset temperature remains low, and it is widely believed that in-gap features are introduced by the transition metal dopant that are detrimental to performance. Therefore, magnetic doped TIs, e.g., V-, Cr- and Mn $Sb_2Te_3$[7-10], Fe- and Mn-doped single crystals of $Bi_2Te_3$[11, 12] and Cr-, Mn-doped of $Bi_2Se_3$[13, 14], are being increasingly studied to determine the optimal set of dopants, magnetic order and transport properties in thin films and single crystals. In this work, we report the electronic effects of doping $Sb_2Te_3$ single crystals with iron studied with magnetotransport experiments and complementary *ab initio* calculations.

$Sb_2Te_3$ has a tetradymite crystal structure, which belongs to the $R\bar{3}m$ space group, with quintuple (Te-Sb-Te-Sb-Te) layers piled up along the *c*-axis via the van de Waals interaction. As the ionic radii of the $Sb^{3+}$ is very similar to a number of transition metals $(R(Sb^{3+}) = 0.76$ Å)[15] it reasonable to expect efficient transition metal doping in this family in contrast to $Bi_2Te_3$ where a larger mismatch occurs. Another potential advantage of $Sb_2Te_3$ is that the chemistry ratio of Sb:Te can be controlled by the growth process. By increasing Te pressure, the typical Te vacancies in single crystal $Sb_2Te_3$ can be reduced, which, remarkably, also depletes the bulk conduction channels, thus reducing the carrier density. In tandem, this increases the carrier mobility, e.g. mobilities as high as 25000 $cm^2V^{-1}s^{-1}$ have been attained, which are the highest for any of the existing topological materials[16]. We note that a past DFT study on the transition-metal doped 3D TIs predicts the Fe dopant is unique amongst the potential transition metal dopants because it is predicted to contribute to the density of states and result in multiple electron or hole pockets at the Fermi level in Fe doped $Sb_2Te_3$.[17] This would be very interesting as the Fe doping is likely to have multiple effects by: 1) introducing a ferromagnetic or paramagnetic state; 2) tuning both types and density of charge carrier; 3) modifying the non-trivial transport state due to the intrinsic strong spin-orbit coupling in the $Sb_2Te_3$. However, to the best of our knowledge, there is an absence of experimental study for the Fe-doped $Sb_2Te_3$ system. This motivated us to investigate the Fe doping effect on the electronic structure of $Sb_2Te_3$ by performing quantum transport measurements.

## 2. Methods

### *2.1. Experimental*

High-quality $Sb_2Te_3$ and $Fe_{0.04}Sb_{1.96}Te_3$ (FST) single crystals were grown by the melting and slow-cooling method. Briefly, high-purity stoichiometric amount (~12 g) of Fe, Sb and Te powder were sealed in a quartz tube as starting materials. The crystal growth was carried out in a vertical furnace using the following temperature procedures: i) Heating the mixed powders to completely melt ii) Cooling rapidly to slight above the melting point, iii) Slowly cooling down to room temperature to crystallize the sample. After growth, single crystal flakes with a typical size of 5 * 5 * 0.2 $mm^3$ could be easily exfoliated mechanically from the ingot. Naturally, single crystals prefer to cleave along the [001] direction, resulting in the normal direction of these flakes being (001), as shown the inset photograph of **Fig. 1(b, c)**. The strong (00*l*) (*l* = 6, 9, 12, 15, 18) diffraction peaks **(Fig. 1(b, c))** indicate that the single crystal is a *c*-oriented crystal flake.

The electronic transport properties were measured by the standard four-probe method using a physical property measurement system (PPMS-14T, Quantum Design). Ohmic contacts were prepared on a fresh cleavage *ab* plane using room-temperature cured silver paste. The electric current is parallel to the hexagonal *ab* plane while the magnetic field is orientated along the c-axis. Angle-dependent magnetoresistance (MR) was also measured using a standard horizontal rotational rig mounted on the PPMS.

### *2.2. Theoretical*

Density Functional Theory (DFT) calculations were carried out using the plane-wave code, Vienna Ab-initio Simulation Package (VASP) version 5.44.[18, 19] The Generalized-Gradient Approximation using the Perdew–Burke–Ernzerhof (PBE) exchange-correlation functional was employed, together with the spin orbit interaction and Hubbard correction for the Fe d levels. The Hubbard terms were taken as U = 3 eV and J = 0.28 to be consistent with recent work, although investigations were performed for a range of values and found to yield similar results. Both pure $Sb_2Te_3$ and $Sb_{2-x}Fe_xTe_3$ were simulated with identical levels of precision. For the $Sb_{2-x}Fe_xTe_3$, a 2x2x1 supercell was constructed containing a single Fe atom within a total of 60 atoms. An energy cut-off and electronic convergence of 300 eV and $1.0 \times 10^{-5}$ eV respectively. Forces were converged within 0.02 eV/A. Dispersion corrections were included via the Grimme D3 method to account for van der Waals interactions[20]. A k-point mesh equivalent to a 16×16×2 mesh in the hexagonal unit cell was adopted. For visualizing the Fermi surface, a grid of spacing 0.1 Å$^{-1}$ was constructed corresponding to 20×20×4 in the unit cell, and values were interpolated between these points.

### 3. Results and Discussion

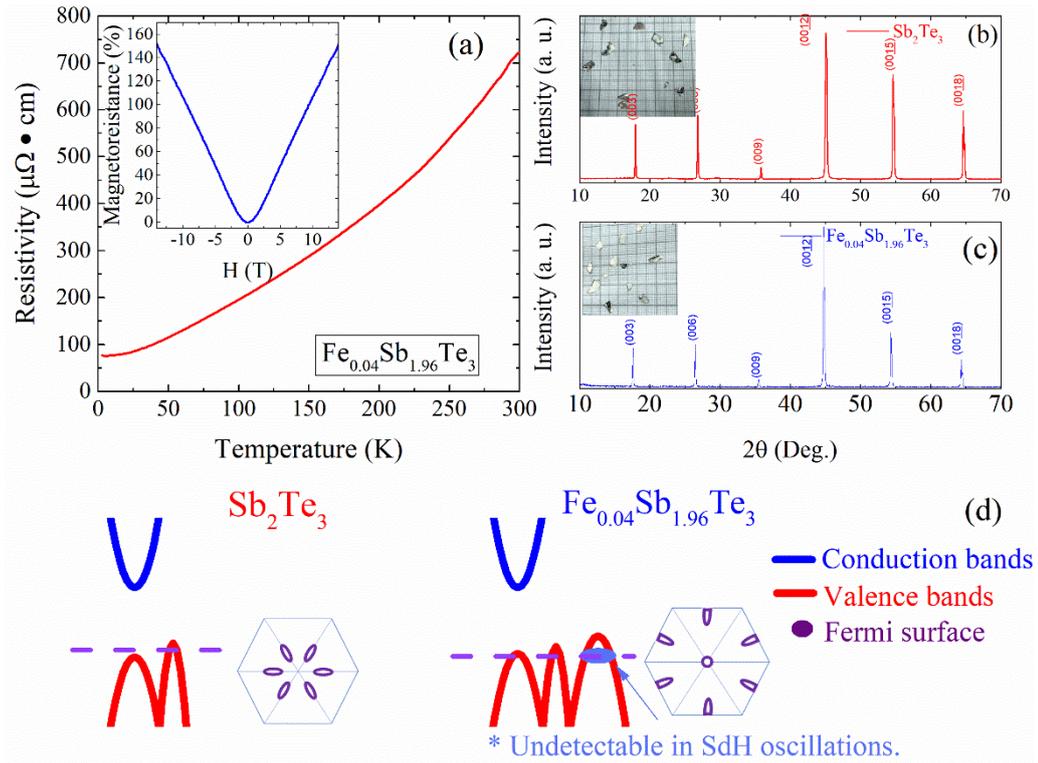

**Fig. 1** (a) Resistivity of $Fe_{0.04}Sb_{1.96}Te_3$ as a function of temperature from 3 to 300 K. The inset shows the MR at 3 K up to 14 T. (b, c) the X-ray diffraction patterns of $Sb_2Te_3$ and FST single crystals along the *c* direction. The inset photos are crystal pieces cleave from the ingot. (d) Sketches of the band structure and Fermi surface of $Sb_2Te_3$ and FST.

Since the bonding between Sb and Te is relatively weak, it can lead to defects (e.g., Te vacancies) in the crystals, and this results in a metallic base state in $Sb_2Te_3$[21]. This situation also occurs in our crystals. As shown in Fig. 2, the resistivity monotonically decreases with the cooling from 300 K to 3 K. At 3 K, the resistivity of FST is about 80 μΩ·cm, presenting a good metallic state. Besides, the residual resistance ratio (RRR = R(300 K)/R(3 K)) is about 10, also indicating the good metallicity. The inset of Fig. 1 shows the magnetoresistance (MR), which defined by (R(H) – R(0))/R(0), of FST single crystal with magnetic field perpendicular to the current. One can see the maximum MR value at 3 K is about 150%, similar to the MR previously studies in this system. Moreover, oscillations patterns could be found in the MR curve, attributed to the Shubnikov-de Haas (SdH) effect. The aforementioned XRD results, as well as the magneto- transport properties, demonstrate the good quality of as-grown single crystals.

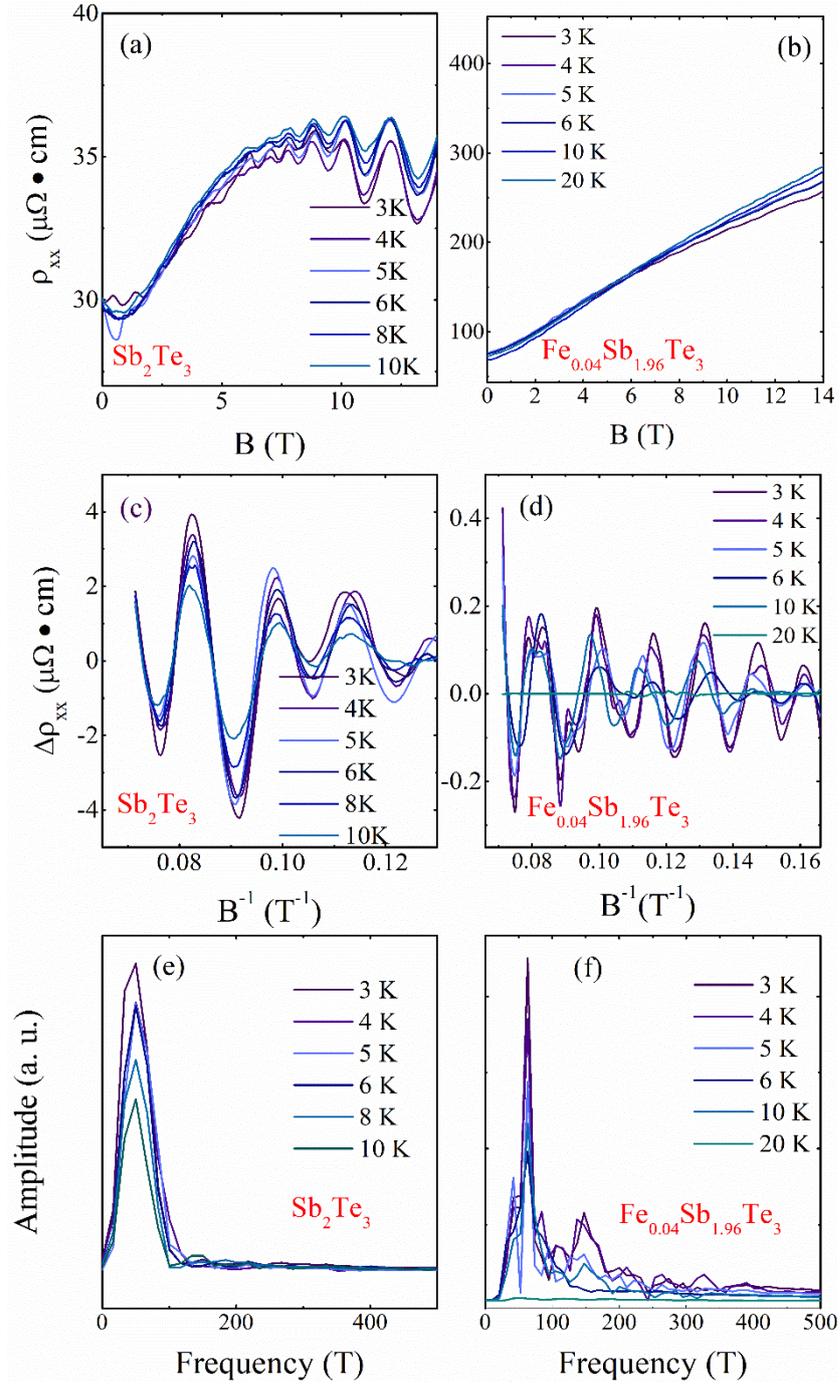

**Fig. 2** SdH oscillations of both Sb₂Te₃ and FST single crystals. (a) and (b), MR curves of Sb₂Te₃ and FST single crystals measured at different temperatures; (c) and (d), oscillation patterns obtained by subtracting the smooth backgrounds at various temperatures, plotted as a function of 1/H; (e) and (f), amplitudes of the the fast Fourier transform from the oscillations.

To further understand the SdH effect in Sb₂Te₃ and Fe-doped single crystals, we conducted a series of MR measurements at different temperatures shown in Fig. 2 (a) and (b). In the pure Sb$_2$Te$_3$ crystals, the MR curves firstly increase with external magnetic fields, however, more-or-less saturate in above 8 T. The Fe-doped samples behaves significantly differently, resulting in a linear-like, unsaturated MR curves within the measured temperature range. The SdH oscillation amplitudes ($\Delta\rho = \rho - \langle\rho\rangle$) are obtained by subtracting a smooth non-oscillatory background $\langle\rho\rangle$ from the oscillating total resistivity ($\rho$). These are plotted against the inverse magnetic field in the Panel (e) and (f) of Fig. 2. The oscillation amplitude displays complex periodic behaviors and decreases with increasing temperature. It is noteworthy that the oscillation amplitude of Sb$_2$Te$_3$ is one order-of-magnitude larger than it in FST, which is taken to imply that there are additional defects induced by the doping process that contribute to charge-carrier scattering. In Panel (c) and (d), one can see the oscillation patterns are different from each other: while there is a single oscillatory mode in Sb$_2$Te$_3$, numerous frequencies appear for the FST. To analyze, the oscillations, we performed fast Fourier transforms as shown in Panel (e) and (f). A single peak at ~50 T can be found in Sb$_2$Te$_3$, indicating that a single pocket exists near the Fermi surface as shown in Fig. 1(d), which agrees well with the band structure calculation [4], and also the previous sdH experiments on p-type Sb$_2$Te$_3$.[22]. Note that, the peak value decreases as the temperature increases. In the present study, multiple oscillation peaks with frequencies at about 62, 150 T, can be detected, representing the multiple large pockets at the Fermi surface shown in Fig. 1(d), which agrees with the DFT calculations, as discussed in a later section.

The oscillations frequency F is related to the cross section of the Fermi surface A, by the Onsager relation: $F = (h/2\pi e)\cdot A$; here, $h$ is Plank's constant, $e$ is the elementary charge. Therefore, the cross section of Sb$_2$Te$_3$ is 0.48 nm$^{-2}$, while in FST, more than one cross section can be detected with an area of 0.62 and 1.44 nm$^{-2}$, respectively. The relative Fermi wave vector $K_F$ can be calculated as $K_F = (A/\pi)^{-2}$ giving 0.29, 0.44 and 0.68 nm$^{-1}$. According to LK theory[23], $M_{os} \propto R_T R_D \sin(\frac{2\pi F}{B} + \beta)$, where $M_{os}$ is the magnitude of oscillation, $R_T$ is the temperature damping factor, $R_D$ is the Dingle damping factor and $\beta$ is the Berry phase, respectively. The effective mass m* can be extracted from the temperature dependence of the SdH oscillation amplitudes by $R_T = \frac{\alpha T m*}{B \cdot \sinh(\alpha T m*/B)}$, in which $\alpha = \frac{2\pi^2 K_B M_e}{eh} \sim 14.96$ T/K, where $k_B$ is the Boltzmann constant and $m_e$ is the electron rest mass. We fit the temperature dependence of SdH oscillation amplitude of $Sb_2Te_3$ and FST as shown in Fig. S2, to obtain the effective masses are 0.23 and 0.37 $m_e$, respectively. The extrapolated Landau-level index ν at the extreme field limit, e.g., 1/B → 0, is related to Berry's phase, which indicates a phase shift regarding the conventional Landau quantization in materials. The intercept ν = 0 corresponds to a normal metal, while ν = 0.5 comes from the massless relativistic fermions in a magnetic field. As shown in Fig. S2, the intercept of $Sb_2Te_3$ and FST single crystal are -0.5 and -0.3 (0.7), respectively, indicating that the quantum oscillations might be contributed by the surface states in $Sb_2Te_3$, but are sensitive to a mixed contribution of surface states and bulk states in FST. The aforementioned parameters are summarized below in Table 1.

| Sample | Frequency (T) | Area (nm$^{-2}$) | Fermi Vector (nm$^{-1}$) | Effective Mass ($m_e$) | Berry Phase (2π) |
|---|---|---|---|---|---|
| $Sb_2Te_3$ | 50 | 0.48 | 0.29 | 0.23 | 0.5 |
| FST | 62 | 0.62 | 0.44 | 0.37 | 0.7 |
| | 150 | 1.44 | 0.68 | | |

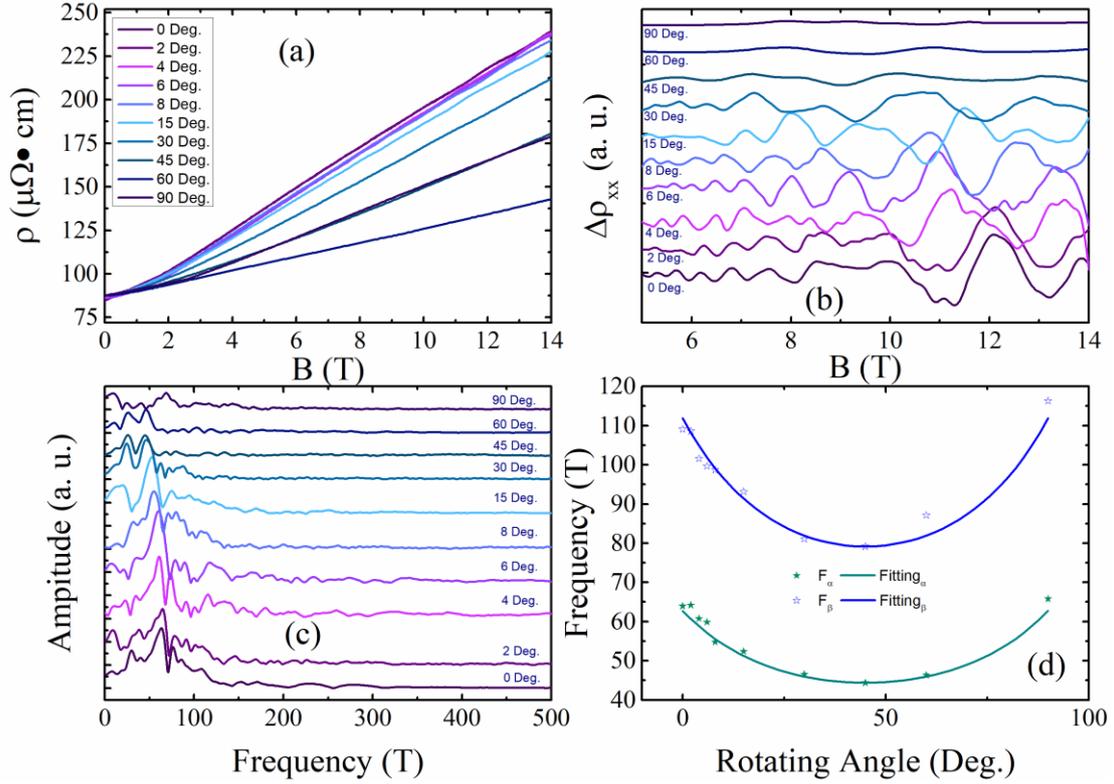

**Fig. 4** Angular-dependence of the SdH oscillations of the FST single crystal. (a) MR curves FST single crystals measured at different angles at 3 K; (b) oscillation patterns obtained by subtracting the smooth backgrounds at various angles, plotted as a function of B; (c), amplitudes plots of fast Fourier transform from the oscillations; (d) two selected frequencies evolve with rotation, in which the star symbols are experimental data, while curves are fitted with $F(\theta)=F(0)/\cos(\theta-45)$.

The angle dependence of quantum oscillations at different tilt angles (θ) provides information about the shape, size and dimensionality of the Fermi surface. Fig. 4 shows the MR at different θ values. In the MR plot displayed in Fig. 4(a), the maximum resistivity value decreases in the rotation process and reaches the minimum value at 60 degrees. Interestingly, the MR value at 90 degrees, which means the external magnetic field is parallel to the current, is larger than that of 60 degrees. and comparable with the magnitude of 45 degrees. During the rotation towards 90 degrees., the oscillation component gradually weakens. After carefully subtracting the background, the oscillation pattern can be found at all positions, which are plotted as a function of H in Fig. 4(b) and 1/B in Fig. 4(c). Obviously, the positions of maxima and minima change systematically with the tilt angle. Fig 4(c) shows the FFT of the quantum oscillations at various angles, showing the multi-frequency nature of FST single crystal at all angles. In Fig. 4(c), it is possible to observe that the frequency $F$ in the rotating process decreases monotonically until 45 degrees., then increases. Since F is proportional to the cross section of the Fermi surface A, we conclude that A is not constant for different magnetic field orientations, which means the Fermi surface is not spherical but ellipsoid or some-other two dimensional like shape. We, therefore, fit the frequencies as a function of rotating angle in form of $F(\theta) = F(45)/\cos(\theta - 45)$. It is noted that the bulk valence band of $Sb_2Te_3$ is confirmed to consist of ellipsoidal valleys which are tilted by 45 degrees from the *c* axis [8]. Therefore, the oscillations may origin from the bulk state, due to the defects during single crystal growth, as well as the Fe doping. Together with the previously DFT calculation [6], we deduce the Fermi surface structure is close to the sketch in the inset of Fig. 1(d).

Moreover, the Landau-level index diagrams of oscillations from 0 to 30 Deg. are plotted in Fig. $S_3$. Due to the multi-frequency nature of those oscillations, we chose one of the main patterns to determine the intercept. Interestingly, all of those intercepts are around -0.3 - -0.5, and are almost unchanging during the rotation process.

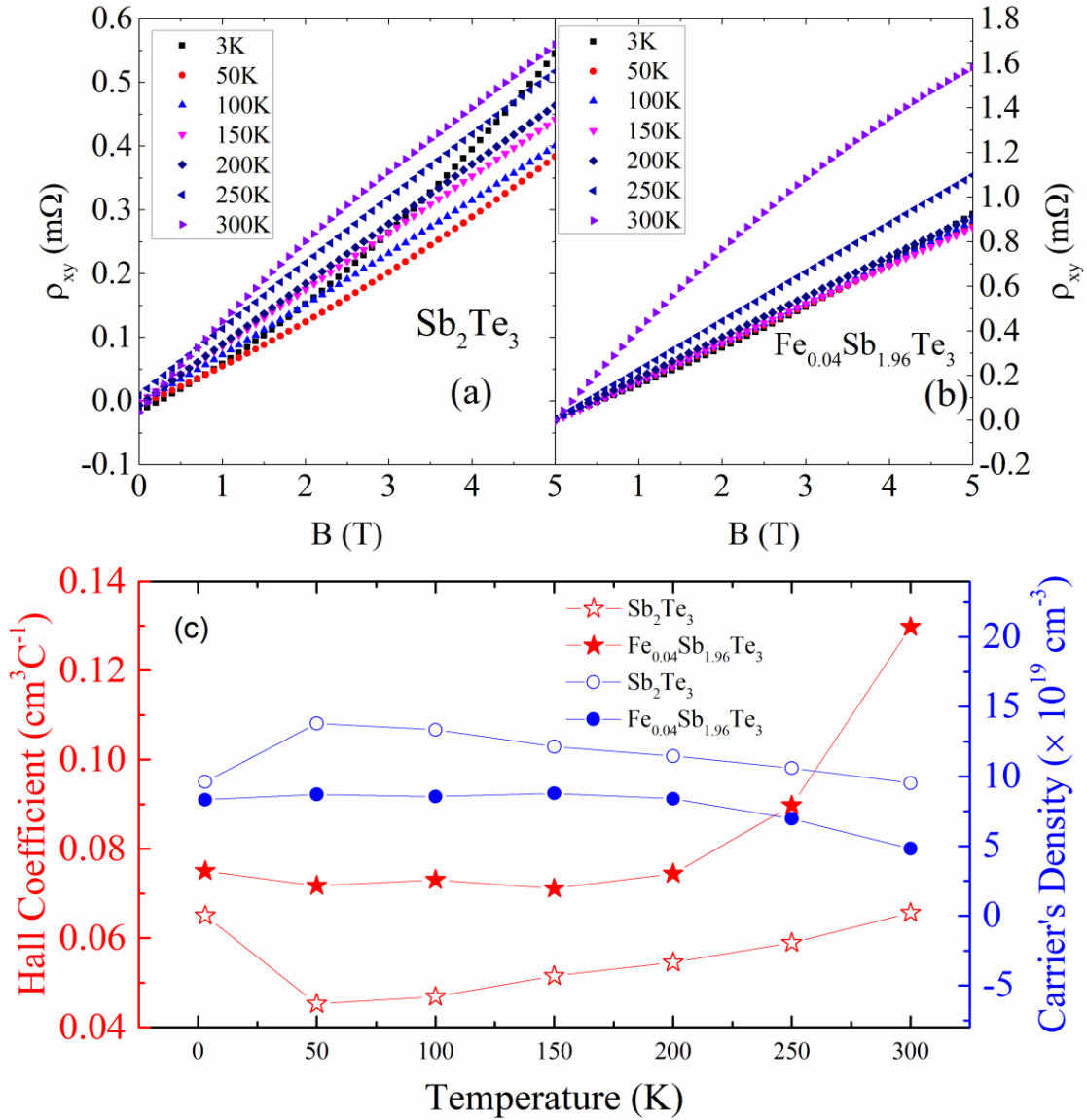

**Fig. 5** Hall Effect of $Sb_2Te_3$ and FST single crystals. (a) and (b), $\rho_{xy}$ plots of $Sb_2Te_3$ and FST single crystals at various temperatures. (c) The Hall coefficient and the carrier density calculated from the relative Hall curves are plotted as functions of temperature.

It is well known that the defects in $Sb_2Te_3$ contribute to the hole carriers, leading to a p-type semimetal. We here introduce Hall measurements to further verify this feature in our as grown $Sb_2Te_3$ single crystal, as well as check the effect of Fe doping. In Fig. 5 (a) and (b), the Hall Effect curves are plotted from 3 to 300 K. One can easily identify that the linear-like increasing curves, which correspond to a p-type conductor, are contributed by hole carriers. Though past DFT calculations indicate that Fe doping may introduce electron carriers into $Sb_2Te_3$ [17], the curves are more likely corresponding to a single-hole carrier model, rather than a multi-carrier model.

At 3 K, the Hall coefficient of $Sb_2Te_3$ and FST are 0.065 and 0.075 $cm^3C^{-1}$, respectively. Together with the conductivity measured before, we can obtain the Hall mobility is 2167 $cm^2$/Vs in $Sb_2Te_3$ and 990 $cm^2$/Vs in FST. Fe-doping slightly reduces the carrier density and mobility of $Sb_2Te_3$, while increasing the defect density. One possible reason is that Fe introduces electron carriers and might reduce the hole carriers contributed in $Sb_2Te_3$. The carrier's density does not change much in the heating process, which might due to the carriers' origin from synthesis defects, not intrinsic carriers which evolves significantly with temperature.

Let us focus on the carrier density and mobility: the Fe dopant slightly reduces the carrier density, and significantly reduces the carrier mobility at 3 K. We attribute the doping effect on carriers to two general aspects: 1) Fe shifts the Fermi level towards the valence band, and thus increases the Fermi pocket area; 2) Fe introduces electron carriers, which might equally reduce the contribution of hole carriers in Hall measurements. Meanwhile, Fe dopant also introduces more defects in the crystal lattice, which may be one of the main reasons that reduce carrier's mobility in FST single crystal. Moreover, Fe dopant changes the Fermi surface morphology to a great extent: 1) more than two Fermi pockets appear in FST, compared to the single pocket in $Sb_2Te_3$; 2) the angle dependence measurements illustrates that those pockets possess a 2D or 2D-like shape. The relative complexity of the Fermi surface offers an an extra free degree of Fermi surface tuning, e.g., via thickness or gating.

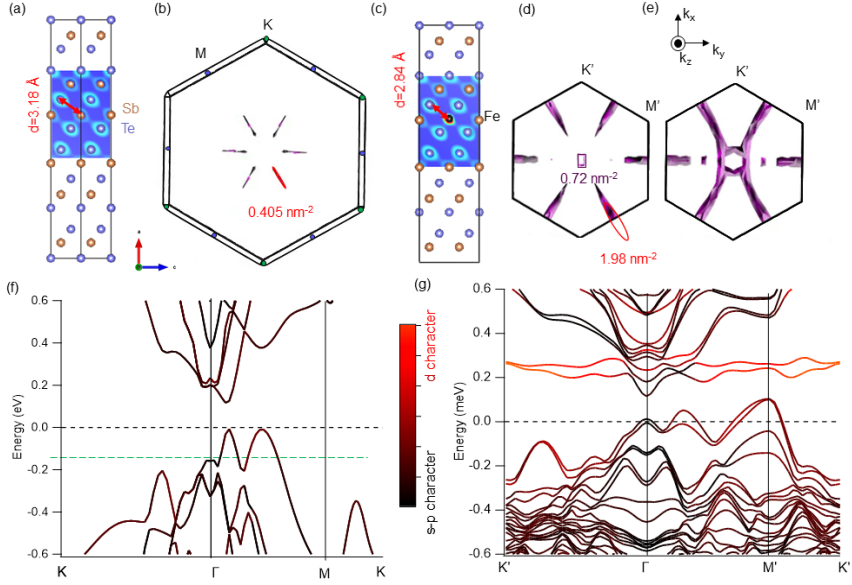

**Fig. 6** DFT calculations for $Sb_2Te_3$ and $Sb_{2-x}Fe_xTe_3$. a) The undoped $Sb_2Te_3$ cell is characterized by a large Sb-Te bond distance d=3.18 Å. The shaded area indicates the valence electron density. b) The Fermi surface calculated for p-type $Sb_2Te_3$ viewed along the (001) projection includes of six major ellipsoids from the hole bands. For comparison, the red ellipsoid depicts an area of 0.405 nm$^{-2}$. c) The Fe dopant (black) causes a crystallographic distortion in the $Sb_2Te_3$ supercell and yields a reduced cation-anion distance d=2.84 Å with an enhanced electron density near the Fe. d) The inner Fermi surface for the $Sb_{2-x}Fe_xTe_3$ includes an additional pocket at the Γ point. The cross-sectional area of the two highlighted features are 0.72 and 1.98 nm$^{-2}$ respectively. e) The exchange-splitting forms a second Fermi surface (right) that wraps the inner $Sb_{2-x}Fe_xTe_3$ surface. e) The band structure of pure $Sb_2Te_3$ indicates a predominant s-p character and shows the light and heavy hole bands at the Fermi surface. The green line is the energy used to draw the Fermi surface.  f) The band-structure of the $Sb_{2-x}Fe_xTe_3$ shows an additional hole band at Γ at the valence band.

To clarify the topology of the Fermi surface features detected in the sDH oscillations, density functional theory calculations were performed. Fig. 6 summarizes the DFT calculations comparing pure $Sb_2Te_3$ with $Sb_{2-x}Fe_xTe_3$. The undistorted crystal structure of $Sb_2Te_3$ in shown in Fig. 6a. The corresponding Fermi surface is shown in Fig. 6 b, where p-type doping has been modelled by shifting the Fermi surface to E = -0.1 eV. The Fermi surface viewed along the $k_z$ projection consists of six ellipsoidal valence bands derived from the hole bands along six-fold symmetric Γ->M lines. This agrees with past work which proposed a six ellipsoid valence band model of $Sb_2Te_3$.[24-26] At moderate doping corresponding to $E_F$=-0.1ev, the maximum cross-sectional area of these light-hole ellipsoids is 0.405 nm$^{-2}$ in reasonable agreement with experimental signature in the sdH oscillations (0.48 nm$^{-2}$). Heavier p-type doping results in a further six heavy-hole pockets becoming populated. The offset between the valence band maxima (VBM) is very sensitive to strain and external pressure in $Sb_2Te_3$.[27, 28]

The Fe dopant introduces a local crystallographic distortion in the $Sb_2Te_3$ cell that is reflected in the band structure. Figure 6 c shows the relaxed chemical structure of a supercell containing a Fe-dopant and indicates that Fe-Te bond-lengths ($d_{Fe-Te}$=2.87 Å) are shorter than the analogous Sb-Te bond-lengths ($d_{Sb-Te}$ = 3.18 Å). Fig. 6 d shows the Fermi surface after Fe-doping. The joint effect of hole-doping and distortion introduced by the Fe modifies the valence band maxima (VBM) features and consequently, the Fermi surface in the $Sb_{2-x}Fe_xTe_3$ has additional features, most prominently at the Γ point. The cross-sectional area of the closed orbits of the ellipsoids is expanded in the Fe-doping to 1.98 nm$^{-2}$ whereas the Γ pocket has a smaller area corresponding to 0.7 nm$^{-2}$. These can be tentatively assigned to the larger and smaller sDH oscillation frequency. Exchange splitting also gives rise to a second larger Fermi surface (Fig. 7 e) which contains open orbitals, making it unlikely to contribute to the sDH oscillations.

The band-structure of the undoped $Sb_2Te_3$ (Figure 6 f) and Fe-doped $Sb_2Te_3$ (Figure 6g) clearly show the contrasting features in the VBM at the Γ and M points which are responsible for the respective Fermi surface. By analyzing the band-character, it is seen that the Γ point in Fe-doped $Sb_2Te_3$ is comprised entirely of Sb-Te s-p orbitals, and therefore hybridization with the Fe d levels does not play a direct role. Instead, this band energy is a reflection of the crystallographic distortion and strain introduced by the dopant.

## 4. Conclusions

With Fe doping, the $Sb_2Te_3$ single crystal still possesses a metallic state with a large MR in low temperature region. Compared with $Sb_2Te_3$, FST shows more complexity in its Fermi surface morphology, which manifests as multi-frequency oscillation patterns in MR measurements. In the angular-dependent measurement, the oscillation frequencies shift, and two of the frequencies indicate 2D-like behavior. We tentatively assign these two features to the two-valence band minimum, one of which is intrinsic to $Sb_2Te_3$, and one which is introduced by chemical strain associated with the Fe dopant. Further, we find that both samples are hole-carrier dominated, and Fe doping reduces the carrier's density, and mobility. These results show that the hybridization of transition-metal defect bands with the intrinsic $Sb_2Te_3$ bands is a crucial consideration if such materials are to be incorporated into future electronic devices.

## 5. Acknowledgements

We acknowledge support from the ARC Professional Future Fellowship (FT130100778), DP130102956, DP170104116, DP170101467and ARC Centre of Excellence in Future Low-Energy Electronics Technologies. This research was undertaken with the assistance of resources and services from the National Computational Infrastructure (NCI), which is supported by the Australian Government.